\documentclass[11pt,a4paper]{article}

\usepackage{graphicx,epsf,epsfig,wrapfig}

\usepackage[latin1]{inputenc}
\usepackage{aeguill} 
\usepackage{natbib,subfigure} 
\usepackage[]{sidecap}
\usepackage{color}
\usepackage{amsmath,amssymb}
\usepackage[T1]{fontenc}
\usepackage{mathptmx}
\usepackage[english]{babel}
\usepackage{lastpage}
\usepackage{lipsum}

\usepackage{geometry}
\usepackage{fancyhdr}    
\usepackage{hyperref}

\hypersetup{
   colorlinks=true,
   citecolor=blue,
   linkcolor=blue,
   urlcolor=black
   }

\def\ssb{\sigma_\mathrm{SB}}

\def\G{G}

\def\msun{ M_\odot}
\def\lsun{ L_\odot}

\def\co2{CO$_2$}
\def\h2o{H$_2$O}
\def\ch4{CH$_4$}
\def\N2{N$_2$}

\def\Rp{R_\mathrm{\,p}}

\def\rhob{\bar{\rho}}
\def\Flux{F}

\def\Ms{M_\star}

\def\Ls{L_\star}

\def\grav{g}
\def\cp{c_\mathrm{p}}

\def\ps{p_\mathrm{s}}

\def\press{p}
\def\Cap{C}
\def\dF{\delta F}
\def\dT{\tilde{\delta T}}
\def\freq{\sigma}
\def\w0{\omega_0}
\def\amp{ q_0}
\def\plm{\tilde{p}_l^m}
\def\ptwotwo{\tilde{p}_2^2}
\def\p22{\tilde{p}_2^2}
\def\pc{\tilde{p}_2^{2*}}
\def\Pquad{\tilde{q}_\mathrm{a}}
\def\ac{a_\mathrm{c}}

\def\PotTh{U_\mathrm{a}}
\def\Ta{T_\mathrm{a}}
\def\Ka{K_\mathrm{a}}
\def\dela{\delta_\mathrm{a}}
\def\phia{\phi_\mathrm{a}}
\def\Tg{T_\mathrm{g}}
\def\Kg{K_\mathrm{g}}
\def\k2{k_2}
\def\bgrav{b_\mathrm{g}}

\def\batm{b_\mathrm{a}}
\def\dt{\Delta t}

\def\mass{m}

\def\tauM{\tau_M}
\def\aM{\alpha}
\def\Al{A_\mathrm{2}}

\def\d{\mathrm{d}}

\def\p{\mathrm{p}}
\def\Im{\mathbb{I}\mathrm{m}}
\def\PLl{P_l}
\def\Ylm{Y_l^m}
\def\P22{P_2^2}
\def\Ylmc{Y_l^{m*}}

\def\vr{\mathbf{r}}
\def\vrp{\mathbf{r'}}

\newcommand{\balign}[1]{
\begin{eqnarray}
#1
\end{eqnarray}}

\newcommand{\eq}[1]{Eq.\,(\ref{#1})}

\newcommand{\fig}[1]{Fig.\,\ref{#1}}

\newcommand{\sect}[1]{Sect.\,\ref{#1}}

\newcommand{\tab}[1]{Table\,\ref{#1}}

\newcommand{\itref}[1]{$\textit{\ref{#1}}$}

\begin{document}
%
%
%


\def\jnl@style{\it}
\def\aaref@jnl#1{{\jnl@style#1}}

\def\aaref@jnl#1{{\jnl@style#1}}

\def\aj{\aaref@jnl{AJ}}                   
\def\araa{\aaref@jnl{ARA\&A}}             
\def\apj{\aaref@jnl{ApJ}}                 
\def\icarus{\aaref@jnl{Icarus}}                 
\def\apjl{\aaref@jnl{ApJ}}                
\def\apjs{\aaref@jnl{ApJS}}               
\def\ao{\aaref@jnl{Appl.~Opt.}}           
\def\apss{\aaref@jnl{Ap\&SS}}             
\def\aap{\aaref@jnl{A\&A}}                
\def\aapr{\aaref@jnl{A\&A~Rev.}}          
\def\aaps{\aaref@jnl{A\&AS}}              
\def\azh{\aaref@jnl{AZh}}                 
\def\baas{\aaref@jnl{BAAS}}               
\def\jrasc{\aaref@jnl{JRASC}}             
\def\memras{\aaref@jnl{MmRAS}}            
\def\mnras{\aaref@jnl{MNRAS}}             
\def\pra{\aaref@jnl{Phys.~Rev.~A}}        
\def\prb{\aaref@jnl{Phys.~Rev.~B}}        
\def\prc{\aaref@jnl{Phys.~Rev.~C}}        
\def\prd{\aaref@jnl{Phys.~Rev.~D}}        
\def\pre{\aaref@jnl{Phys.~Rev.~E}}        
\def\prl{\aaref@jnl{Phys.~Rev.~Lett.}}    
\def\pasp{\aaref@jnl{PASP}}               
\def\pasj{\aaref@jnl{PASJ}}               
\def\qjras{\aaref@jnl{QJRAS}}             
\def\skytel{\aaref@jnl{S\&T}}             
\def\solphys{\aaref@jnl{Sol.~Phys.}}      
\def\sovast{\aaref@jnl{Soviet~Ast.}}      
\def\ssr{\aaref@jnl{Space~Sci.~Rev.}}     
\def\zap{\aaref@jnl{ZAp}}                 
\def\nat{\aaref@jnl{Nature}}              
\def\iaucirc{\aaref@jnl{IAU~Circ.}}       
\def\aplett{\aaref@jnl{Astrophys.~Lett.}} 
\def\apspr{\aaref@jnl{Astrophys.~Space~Phys.~Res.}}
\def\bain{\aaref@jnl{Bull.~Astron.~Inst.~Netherlands}} 
\def\fcp{\aaref@jnl{Fund.~Cosmic~Phys.}}  
\def\gca{\aaref@jnl{Geochim.~Cosmochim.~Acta}}   
\def\grl{\aaref@jnl{Geophys.~Res.~Lett.}} 
\def\jcp{\aaref@jnl{J.~Chem.~Phys.}}      
\def\jgr{\aaref@jnl{J.~Geophys.~Res.}}    
\def\jqsrt{\aaref@jnl{J.~Quant.~Spec.~Radiat.~Transf.}}
\def\memsai{\aaref@jnl{Mem.~Soc.~Astron.~Italiana}}
\def\nphysa{\aaref@jnl{Nucl.~Phys.~A}}   
\def\physrep{\aaref@jnl{Phys.~Rep.}}   
\def\physscr{\aaref@jnl{Phys.~Scr}}   
\def\planss{\aaref@jnl{Planet.~Space~Sci.}}   
\def\procspie{\aaref@jnl{Proc.~SPIE}}   

\let\astap=\aap
\let\apjlett=\apjl
\let\apjsupp=\apjs
\let\applopt=\ao

\begin{center}
\vspace{10pt}


\vspace*{40pt}

 \Large{\textbf{Asynchronous rotation of Earth-mass planets\\ in the habitable zone of lower-mass stars}}

\vspace{30pt}

 \large{Jérémy Leconte$^{1,2,3,4}$, Hanbo Wu$^{1,5}$, Kristen Menou$^{3,6}$, Norman Murray$^{1,5}$}

\vspace{25pt}
$^{1}$Canadian Institute for Theoretical Astrophysics,
\\ 
60st St George Street, University of Toronto, Toronto, ON, M5S3H8, Canada\\
$^{2}$Banting Fellow\\
$^{3}$ Center for Planetary Sciences, Department of Physical \& Environmental Sciences,\\
 University of Toronto Scarborough, Toronto, ON, M1C 1A4\\
 $^{4}$Laboratoire de M\'et\'eorologie Dynamique, Institut Pierre Simon Laplace, 4 Place Jussieu, BP 99, 75252 Paris, France.\\ 
 $^{5}$Department of Astronomy \& Astrophysics,
 University of Toronto, Toronto, ON, M5S 3H8\\
 $^{6}$ Department of Physics, University of Toronto, 60 St George Street, Toronto, ON M5S 1A7\\
\vspace{5pt}
E-mail: jleconte@cita.utoronto.ca
\end{center}

\vspace{25pt}

\vspace{25pt}

\begin{center}
\begin{minipage}{13cm}
\textbf{Planets in the habitable zone of lower-mass stars are often assumed to be in a state of tidally synchronized rotation, which would considerably affect their putative habitability. Although thermal tides cause Venus to rotate retrogradely, simple scaling arguments tend to attribute this peculiarity to the massive Venusian atmosphere. Using a global climate model, we show that even a relatively thin atmosphere can drive terrestrial planets' rotation away from synchronicity. We derive a more realistic atmospheric tide model that predicts four asynchronous equilibrium spin states, two being stable, when the amplitude of the thermal tide exceeds a threshold that is met for habitable Earth-like planets with a 1\,bar atmosphere around stars more massive than $\sim0.5-0.7\,\msun$. 
Thus, many recently discovered terrestrial planets could exhibit asynchronous spin-orbit rotation, even with a thin atmosphere.
}
\end{minipage}
\end{center}

\vspace{35pt}

As we all experience in our everyday life, atmospheric temperatures oscillate following the diurnal insolation cycle. This, in turn, creates periodic large-scale mass redistribution inside the atmosphere, the so-called thermal atmospheric tides. But as we all have experienced too, the hottest moment of the day is actually not when the Sun is directly overhead, but a few hours later. This is due to the thermal inertia of the ground and atmosphere that creates a delay between the solar heating and thermal response (driving mass redistribution), causing the whole atmospheric response to lag behind the Sun (\itref{CL70}).

Because of this asymmetry in the atmospheric mass redistribution with respect to the sub-solar point, the gravitational pull exerted by the Sun on the atmosphere has a non-zero net torque that tends to accelerate or decelerate its rotation, depending on the direction of the solar motion (\itref{GS69}, \itref{ID78}). Because the atmosphere and the surface are usually well coupled by friction in the atmospheric boundary layer, the angular momentum transferred from the orbit to the atmosphere is then transferred to the bulk of the planet, modifying its spin (\itref{DI80}).

On Earth, this effect is negligible because we are too far away from the Sun, but the atmospheric torque due to thermal tides can be very powerful, as seen on Venus. Indeed, although tidal friction inside the planet is continuously trying to spin it down to a state of synchronous rotation, thermal tides are strong enough to drive the planet out of synchronicity and to force the slow retrograde rotation that we see today (\itref{GS69}-\itref{CL03b}). Very simple scaling arguments predict that the amplitude of the thermal tide is proportional to the ratio of the atmospheric mean surface pressure over its scale height (\itref{CL70}). Everything else being equal, one would thus expect the thermal tide to be $\sim50$ times weaker if Venus had a less massive, cooler Earth-like atmosphere. Does this scaling really hold and how massive does an atmosphere need to be to affect the planetary rotation, this has not been completely worked out yet.

These questions are of utmost importance as we now find many terrestrial planets in a situation similar to Venus. Because the habitable zone is closer around lower-mass stars, planets in this region are often expected to be tidally synchronized with the orbit (\itref{KWR93}-\itref{LFC13}), which seems to create additional difficulties for their habitability. In particular, the permanent night side can be an efficient cold-trap for water (\itref{Men13}, \itref{LFC13}), strongly destabilize the carbonate-silicate cycle (\itref{KGM11}, \itref{EKP12}), and even cause atmospheric collapse in extreme cases (\itref{JHR93}, \itref{HK12}).


 \begin{SCfigure}[1.3][hbtp]
  \centering
    \includegraphics[width=0.5\textwidth]{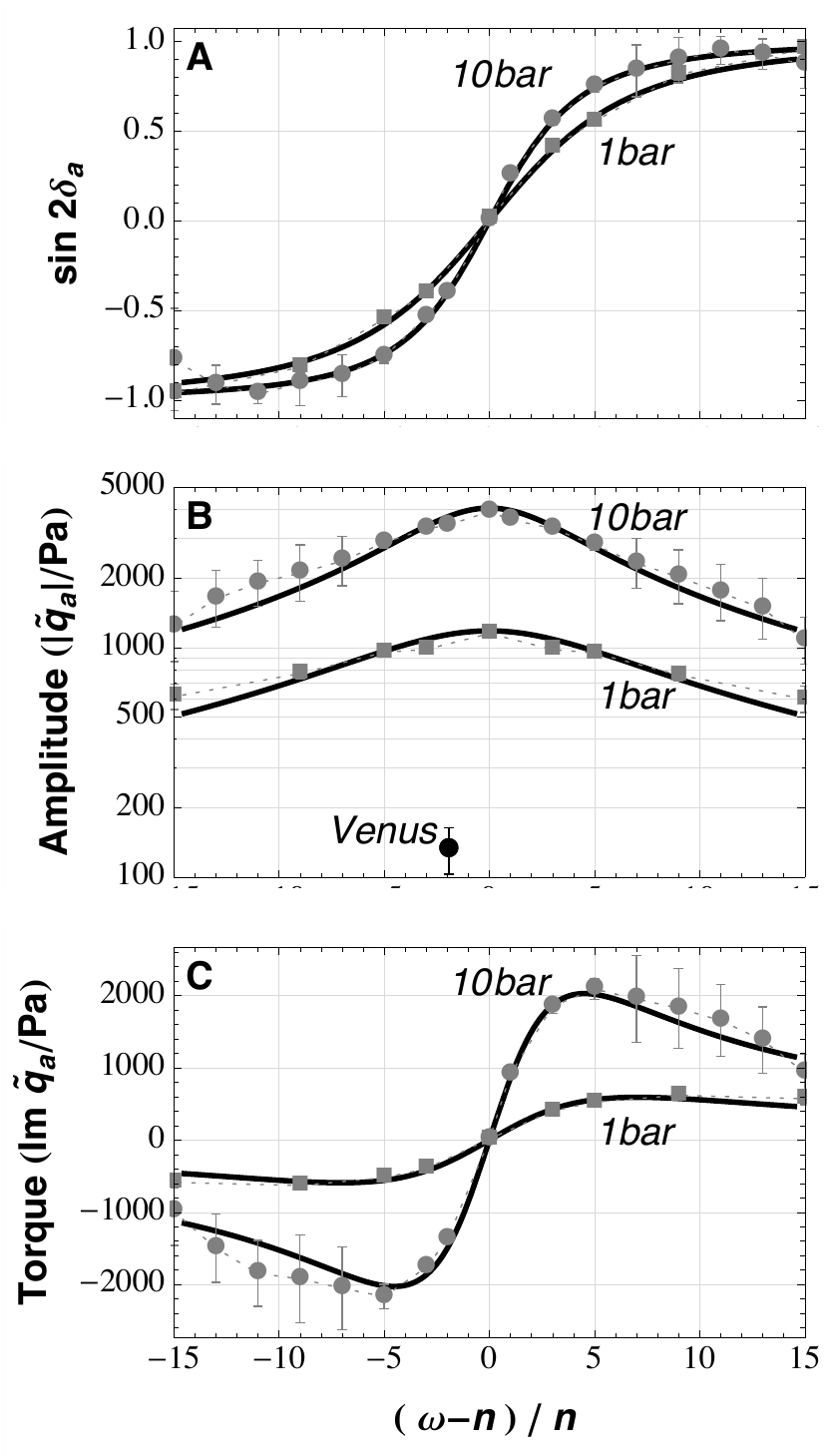}
    \caption{\textbf{Frequency dependence of the torque.} Sin of the lag angle ($\sin 2\dela$; \textbf{A}), amplitude of the pressure bulge ($|\Pquad|$; \textbf{B}), and torque ($\Im(\Pquad)$; \textbf{C}) as a function of the forcing frequency ($(\omega-n)/n$) computed from the numerical atmospheric model (gray data points) and given by our analytical model (black curve). Results are shown for two pressures (1 and 10 bar) in the 1366\,W/m$^{-2}$ case with an orbital period of 225\,d. The error bars show the internal variability ($\pm1\sigma$). The Venusian tide amplitude is shown in \textbf{B}. Despite its simplicity, the analytical model fairly captures the frequency dependence of the thermal tide response.}
      \label{fig:analytical_fit}
\end{SCfigure}
 
Here, we investigate whether or not thermal tides can drive terrestrial planets in the habitable zone, with a relatively thin atmosphere, out of synchronous rotation. Previous studies on Venus (\itref{GS69}-\itref{CL03b}), and for exoplanets (\itref{CLL08}, \itref{CCL14}), have shown that this reduces to the search for equilibrium rotation states for which the bodily torque ($\Tg$) and the atmospheric torque ($\Ta$) cancel each other and provide a restoring force against deviations from this equilibrium. In the circular case with zero obliquity, these torques are given by (\itref{CL01}-\itref{CL03b})
 \balign{
 \label{torques}
 \Ta=-\frac{3}{2}\,\Ka \,\batm\left(2\omega-2n\right)\,\,\mathrm{and}\,\, \Tg=-\frac{3}{2}\, \Kg \, \bgrav\left(2\omega-2n\right),
 }
where $\Ka\equiv (3 \,\Ms \Rp^3)/ (5\rhob a^3)$, $\Kg\equiv G \Ms^2 \Rp^5/a^6$, $G$ is the universal gravitational constant, $\Ms$ is the stellar mass, $\Rp$, $\omega$, and $\rhob$ the planet's radius, rotation rate, and mean density, and $a$ and $n$ its orbital semi-major axis and mean motion. $\bgrav(\sigma)$ characterizes the frequency-($\sigma$-)dependent response of the body of the planet (i.e. its rheology), and $\batm(\sigma)$ that of the atmosphere.
 Note that, unlike trapping in asynchronous spin-orbit resonances (like Mercury; \itref{MBE12}), thermal tides do not need any eccentricity to drive a planet out of synchronous rotation. 
While some general conclusions can be reached without a precise knowledge of these responses (\itref{CL01}, \itref{CL03b}, \itref{CLL08}, \itref{CCL14}), the biggest limitation in predicting the properties of equilibrium spin states lies in the uncertainties on the shape and amplitude of $\batm(\sigma)$. In particular, the relation between the mass of the atmosphere and the strength of the thermal tide, the key ingredient to quantify whether asynchronicity is ubiquitous, remains unknown.

To tackle this issue, we use a generic global climate model (\itref{LFC13}, \itref{FWM13}, \itref{LFC13b}), commonly used to model planets in the habitable zone of low-mass stars, to empirically quantify the amplitude of torque induced by thermal tides for planets with various atmospheric masses (characterized by the surface pressure, $\ps$), compositions, and incoming stellar fluxes ($\Flux$). Once $\ps$ and $\Flux$ are chosen, we run the atmospheric model for several diurnal frequencies, $\freq\equiv \omega-n$. Because of the thermal forcing, a surface pressure pattern lagging behind the substellar point forms (see Fig.\,S1). Once mean thermal equilibrium is reached, we compute the complex amplitude of the quadrupolar thermal tide ($\Pquad$; \itref{methods}), as shown in \fig{fig:analytical_fit}, and the value of the torque is given by equation (\itref{torques}) where $\batm(2\omega-2n)=-\sqrt{\frac{10}{3\pi}}\,\Im( \Pquad(\omega-n)).$

To test our framework, we applied our model to the Earth and Venus, for which results meet existing constraints (see Fig.\,S2; \itref{methods}). However, rather counterintuitively, for the same forcing frequency, the amplitude of the thermal tide in a 1\,bar Earth-like atmosphere is almost an order of magnitude stronger than on Venus (\fig{fig:analytical_fit}.B). This difference is the result of the sunlight being almost completely scattered or absorbed before it reaches Venus' surface (\itref{methods}).

 As discussed earlier, the lag and amplitude of thermal tides are closely related to the thermal inertia of the system. In fact, a careful analysis of the result of the climate model shows that the frequency dependence of the atmospheric response shown in \fig{fig:analytical_fit} is fairly analogous to the thermal response of a radiating slab with a finite thermal inertia that is periodically heated, so that, to a very good approximation, we can write 
\balign{
\Pquad(\freq)=-\frac{\amp}{1+i \freq/\w0},
}
  where $i^2=-1$, $\w0$ is the inverse of the timescale needed for the system to reach thermal equilibrium, and $\amp$ is the amplitude of the quadrupole term of the pressure field at zero frequency (\itref{methods}). In theory, $\w0$ can be estimated if the heat capacity of the system is known (\itref{methods}), but in practice, both $\w0$ and $\amp$ are computed from the numerical model for a given atmosphere and are shown in \tab{tab:num_results} for limit cases. In the circular case with zero obliquity, the torque thus writes,
\balign{
\Ta(\omega-n,\ps,\Flux)=\frac{3}{2}\,\Ka \,\amp(\ps,\Flux)\ \frac{(\omega-n)/\w0(\ps,\Flux)}{1+\left[(\omega-n)/\w0(\ps,\Flux)\right]^2}.
}
 Interestingly, although governed by different physics, the atmospheric torque follows the same law as the body torque for a viscoelastic sphere (with the opposite sign).

\begin{figure}[b] 
 \centering
\subfigure{ \includegraphics[scale=.99,trim = 0cm .9cm 0.cm 0.cm, clip]{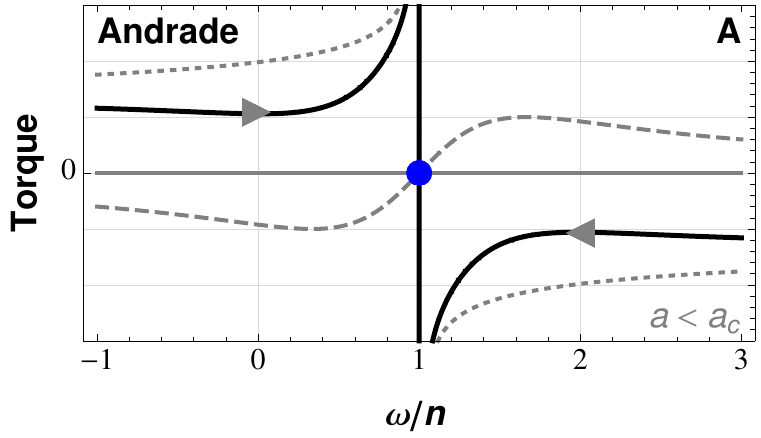} }
\subfigure{ \includegraphics[scale=.99,trim = 0.8cm .9cm 0.cm 0.cm, clip]{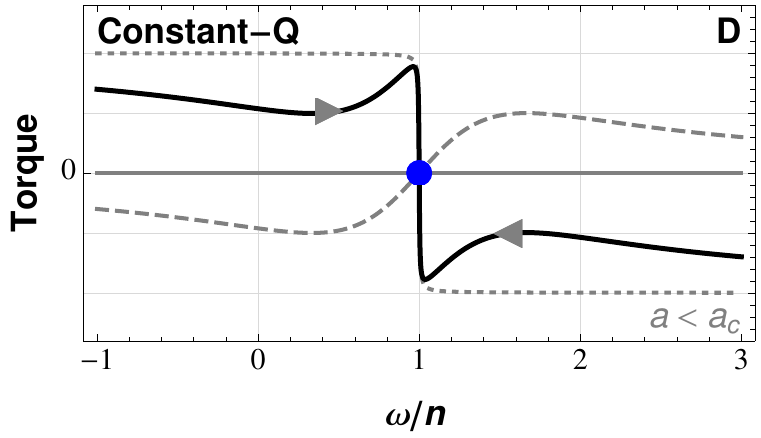} }\\
\vspace{-0.4cm}
\subfigure{ \includegraphics[scale=.99,trim = 0cm .9cm .0cm 0cm, clip]{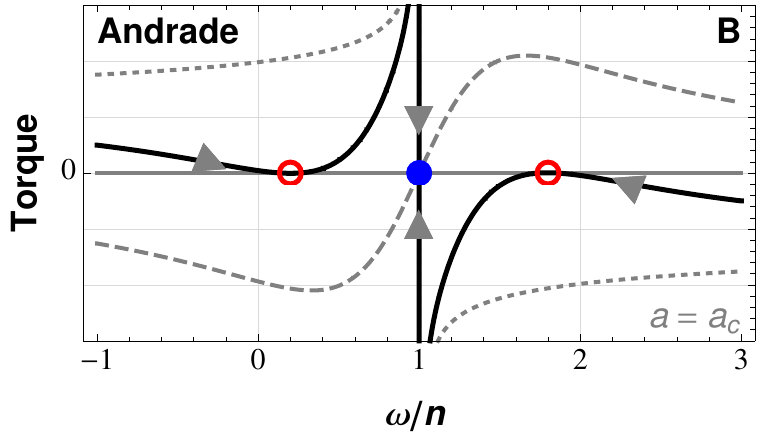} }
\subfigure{ \includegraphics[scale=.99,trim = 0.8cm .9cm 0.cm 0.cm, clip]{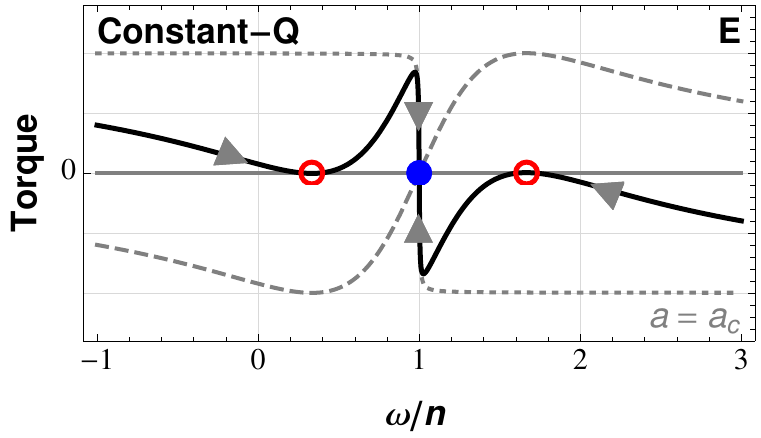} }\\
\vspace{-0.4cm}
\subfigure{ \includegraphics[scale=.99,trim = 0cm .cm .0cm 0cm, clip]{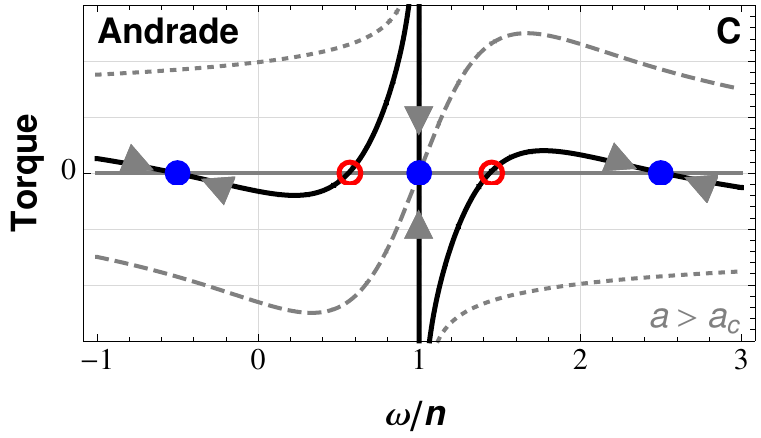} }
\subfigure{ \includegraphics[scale=.99,trim = 0.8cm .cm 0.cm 0.cm, clip]{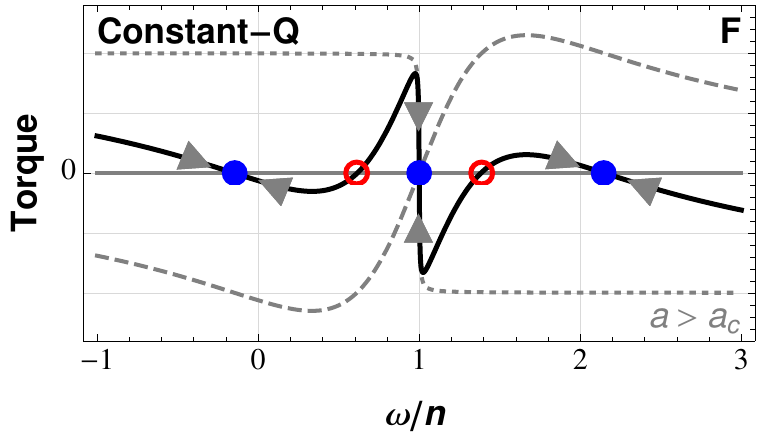} }
\caption{
\textbf{Equilibrium spin states of the planet.} Atmospheric (dashed), gravitational (dotted) and total (solid) torque as a function of spin rate for two tidal models (Andrade: \textbf{A}, \textbf{B}, \textbf{C}; Constant-$Q$: \textbf{D}, \textbf{E}, \textbf{F}).  Arrows show the sense of spin evolution. \textbf{A, D}: weak atmospheric torque, only one equilibrium, synchronous spin state exists (blue disk). \textbf{B, E}: bifurcation point ($a=\ac$). \textbf{C, F}: the atmospheric torque is strong enough to generate four asynchronous, equilibrium spin states, two being \textit{unstable} (red circles) and two being \textit{stable} (blue disks; one is retrograde in the case shown). The synchronous spin state remains stable.
}
 \label{fig:torque}
\end{figure}

A first \textit{qualitative} difference with previous works (\itref{ID78}-\itref{CL01}, \itref{CLL08}, \itref{CCL14}), is that we derive a very different functional form for the atmospheric torque. In particular, the function $f(\freq)\equiv \batm(2\freq)/\bgrav(2\freq)$ (\itref{CL01}) is \textit{not} monotonic around potential equilibria when a realistic rheology is used. As seen in \fig{fig:torque}, for a constant-$Q$ or an Andrade rheology, this results in the possible existence of up to five equilibria in the circular case, two of them being unstable (\itref{methods}). The diversity of equilibria might be even richer in eccentric systems where these numbers could change (\itref{CLL08}, \itref{CCL14}). Interestingly, the synchronous spin state is stable. Knowing that Venus, despite such a rheology, did not end up synchronized tells us that a planet can avoid being trapped in such a stable synchronous state and constrains the history of the Venusian atmosphere (\itref{methods}).

\begin{table}[t]
\centering
\caption{\small Numerical values of the amplitude of the atmospheric quadrupole ($\amp$) and intrinsic thermal frequency of the atmosphere ($\w0$) derived from the global climate model.} \label{tab:num_results}
{\small
\begin{tabular}{@{\vrule height 10.5pt depth4pt  width0pt}lrrrrrr}
&&&\multicolumn3c{Model output}\\
\noalign{\vskip-11pt}
Sets of\\
\cline{4-6}
\vrule depth 6pt width 0pt simulations& $\Flux\,(W.$m$^{-2}$)&$\ps$\,(bar)&$\amp$\,(Pa)& $\w0$\,(s$^{-1}$)&$2\pi/\w0$\,(d)\\
\hline
Venus&2610&92&201&3.77$\times10^{-7}$&193\\
&&&&\\
Inner&1366&1&1180&2.30$\times10^{-6}$&32\\
habitable zone&&10&4050&1.46$\times10^{-6}$&50\\
Outer&450&1&890&1.18$\times10^{-6}$&62\\
habitable zone (N$_2$)&&10&2960&7.17$\times10^{-7}$&101\\
Outer&&&&&\\
habitable zone (CO$_2$)&450&10&2590&9.7$\times10^{-7}$&70\\
\hline
\end{tabular}
}
\end{table}

\clearpage

\begin{SCfigure}[1.3][htb] 
 \includegraphics[scale=1.,trim = 0cm 0.cm 0.cm 0.cm, clip]{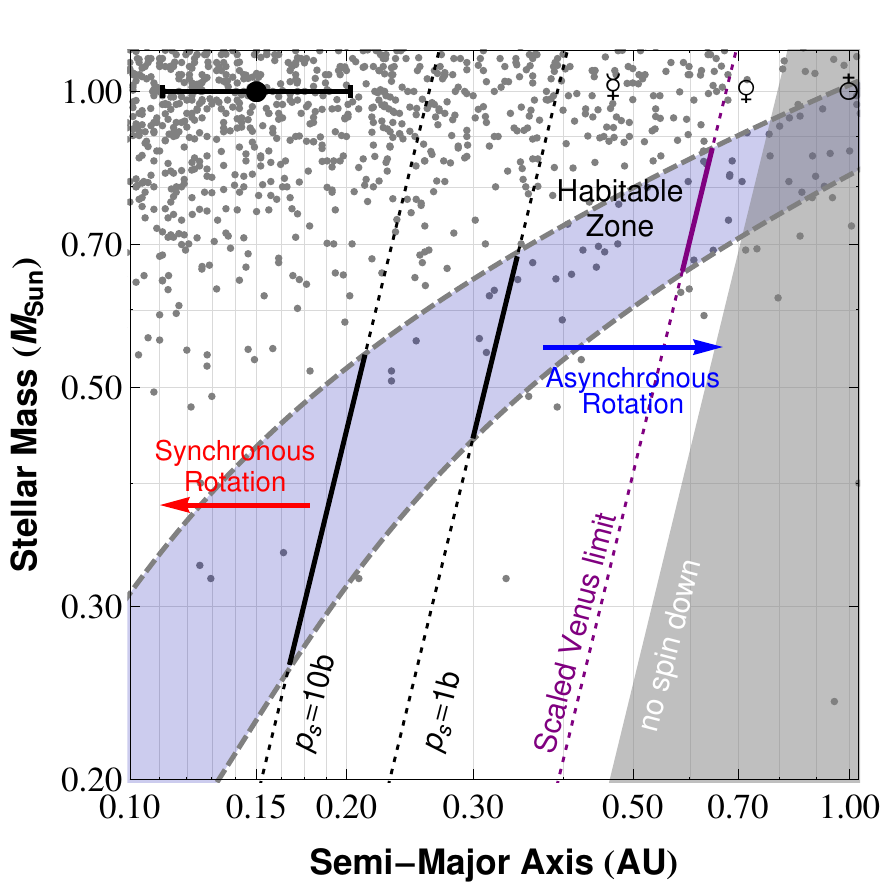} 
\caption{
\textbf{Spin state of planets in the habitable zone.}
Each solid black line marks the critical orbital distance ($\ac$; \eq{critdist}) separating synchronous (on the left, red arrows) from asynchronous planets (on the right, blue arrows) for $\ps=1$ and 10\,bar (the extrapolation outside the habitable zone is shown with dotted lines). Objects in the gray area are not spun down by tides. The habitable zone is depicted by the blue region (\itref{LFC13}, \itref{KRK13}). Gray dots are detected and candidate exoplanets. The error bar illustrates how limits would shift when varying the dissipation inside the planet ($Q\sim100$; \itref{methods}) within an order of magnitude.
}
 \label{fig:threshold}
\end{SCfigure}

In addition, the number and location of equilibria undergoes a \textit{bifurcation} as asynchronous spin states exist \textit{only} when the amplitude of the thermal tide reaches a threshold. Thus, our results reveal the existence a critical distance ($\ac$) beyond which the planet can be asynchronous which, using a constant-$Q$ rheology, reads
\balign{
\label{critdist}
 \ac=\left(\frac{10\pi}{3}\right)^{1/6}\left(G  \Ms  \rhob\Rp^2\,\frac{\k2}{\amp Q}\right)^{1/3},
}
where $\k2$ is the Love number and $Q$ the tidal quality factor (\itref{methods}). Both $\ac$ (\fig{fig:threshold}) and the equilibrium asynchronicity ($|\omega-n|=\w0\left((a/\ac)^3+\sqrt{(a/\ac)^6-1}\right)$; Fig.\,S3) can be computed for various cases using \tab{tab:num_results}.
The corollary is that, even without any triaxiality, planets on circular orbits for which atmospheric tides are too weak should be in exact spin-orbit resonance.

More importantly, our results provide a robust framework for the \textit{quantitative} assessment of the efficiency of thermal tides for different atmospheric masses without having to rely on scaling arguments calibrated on Venus. This is crucial as Venus thermal tides turn out to be relatively weak (see \fig{fig:analytical_fit}.B).
As can be seen in \fig{fig:threshold}, Earth-like planets with a 1\,bar atmosphere are expected to have a non-synchronous rotation if they are in the habitable zone of stars more massive than $\sim$0.5 to 0.7\,$\msun$ (depending on their location in the habitable zone). This lower limit becomes $\lesssim0.3\,\msun$ for a 10\,bar atmosphere.
Interestingly, these limits are much less restrictive than the one obtained from our Venus model (purple line in \fig{fig:threshold}). This realization required full atmospheric modeling.

Atmosphere as thin as 1\,bar being a reasonable expectation value given existing models and Solar System examples, our results demonstrate that asynchronism mediated by thermal tides should affect an important fraction of planets in the habitable zone of lower mass stars. This is especially true in the outer habitable zone where planets are expected to build massive CO$_2$ atmospheres (\itref{KWR93}).

This has many implications. On one hand, all the threats to habitability created by the presence of a permanent cold, night side may not be as severe as usually thought. On the other hand, the habitable zone has been recently shown to be more extended for synchronous planets (\itref{YCA13}). For these objects, if the atmosphere is thick enough, the non-synchronous rotation that will ensue will thus come to limit the extent of the habitable zone around lower-mass stars.

The thermal tide mechanism presented here does not only affect habitable planets, so that many other terrestrial bodies with significant atmospheres could potentially have asynchronous rotations, depending on their orbital location (see \fig{fig:threshold}). With that in mind, observational methods able to constrain the rotation rate of exoplanets (\itref{SMH13}, \itref{SBD14}) become more valuable, and could even be used to constrain the atmospheres of the latter.

\section*{ References and Notes}

\begin{enumerate}

\item \label{CL70} S. Chapman, R. Lindzen, \textit{Atmospheric Tides. Thermal and Gravitational}. Dordrecht: Reidel (1970).

\item \label{GS69} T. Gold, S. Soter, Atmospheric Tides and the Resonant Rotation of Venus. \textit{Icarus}, \textbf{11}, 356-366 (1969).

\item \label{ID78} A. P. Ingersoll, A. R. Dobrovolskis, Venus' rotation and atmospheric tides. \textit{Nature}, \textbf{275}, 37 (1978).

\item \label{DI80} A. R. Dobrovolskis, A. P. Ingersoll, Atmospheric tides and the rotation of Venus. I - Tidal theory and the balance of torques. \textit{Icarus}, \textbf{41}, 1-17 (1980).

\item \label{CL01} A. C. Correia, J. Laskar, The four final rotation states of Venus. \textit{Nature}, \textbf{411}, 767-770 (2001).

\item \label{CL03b} A. C. Correia, J. Laskar, Different tidal torques on a planet with a dense atmosphere and consequences to the spin dynamics. \textit{J. Geophys. Res. (Planets)}, \textbf{108}, 5123 (2003).

\item \label{KWR93} J. F. Kasting, D. P. Whitmire, R. T. Reynolds, Habitable Zones around Main Sequence Stars. \textit{Icarus}, \textbf{101}, 108-128 (1993).

\item \label{HLB11} R. Heller, J. Leconte, R. Barnes, Tidal obliquity evolution of potentially habitable planets. \textit{Astron. Astrophys.}, \textbf{528}, A27 (2011).

\item \label{JHR93} M. M. Joshi, R. M. Haberle, T. T. Reynolds, Simulations of the Atmospheres of Synchronously Rotating Terrestrial Planets Orbiting M Dwarfs: Conditions for Atmospheric Collapse and the Implications for Habitability. \textit{Icarus}, \textbf{101}, 108-128 (1993).

\item \label{KGM11} E. S. Kite, E. Gaidos, M. Manga, Climate Instability on Tidally Locked Exoplanets. \textit{Astrophys. J.}, \textbf{743}, 41 (2011).

\item \label{EKP12} A. R. Edson, J. F. Kasting, D. Pollard, S. Lee, P. R. Bannon, The Carbonate-Silicate Cycle and CO2/Climate Feedbacks on Tidally Locked Terrestrial Planets. \textit{Astrobiology}, \textbf{12}, 562-571 (2012)

\item \label{YCA13} J. Yang, N. B. Cowan, D. S. Abbot, Stabilizing Cloud Feedback Dramatically Expands the Habitable Zone of Tidally Locked Planets. \textit{Astrophys. J.}, \textbf{771}, L45 (2013).

\item \label{Men13} K. Menou, Water-trapped Worlds. \textit{Astrophys. J.}, \textbf{774}, 51 (2013).

\item \label{LFC13} J. Leconte, \textit{et al.}, 3D climate modeling of close-in land planets: Circulation patterns, climate moist bistability, and habitability. \textit{Astron. Astrophys.}, \textbf{554}, A69 (2013).

\item \label{HK12} K. Heng, P. Kopparla, On the Stability of Super-Earth Atmospheres. \textit{Astrophys. J.}, \textbf{754}, 60 (2013).

\item \label{CLL08} A. C. Correia, B. Levrard, J. Laskar, On the equilibrium rotation of Earth-like extra-solar planets. \textit{Astron. Astrophys.}, \textbf{488}, L63-L66 (2008).

\item \label{CCL14} D. Cunha, A. C. Correia, J. Laskar, Spin evolution of Earth-sized exoplanets, including atmospheric tides and core-mantle friction. \textit{Submitted to Astron. Astrophys., Arxiv}, (2014).

\item \label{MBE12} V. V. Makarov, C. Berghea, M. Efroimsky, Dynamical evolution and spin-orbit resonances of potentially habitable exoplanets. The case of GJ 581d. \textit{Astrophys. J.}, \textbf{761}, 83 (2012).

\item \label{FWM13} F. Forget, \textit{et al.}, 3D modelling of the early martian climate under a denser CO2 atmosphere: Temperatures and CO2 ice clouds. \textit{Icarus}, \textbf{222}, 81-99 (2013).
\item \label{LFC13b} J. Leconte, F. Forget, B. Charnay, R. D. Wordsworth, A. Pottier, Increased insolation threshold for runaway greenhouse processes on Earth like planets. \textit{Nature}, \textbf{504}, 268-271 (2013).

\item \label{methods} Supplementary materials are available on Science Online.

\item \label{SMH13} F. Selsis, \textit{et al.}, The effect of rotation and tidal heating on the thermal lightcurves of super Mercuries. \textit{Astron. Astrophys.}, \textbf{555}, A51 (2013).

\item \label{SBD14} I., A., G. Snellen, \textit{et al.}, Fast spin of the young extrasolar planet $\beta$ Pictoris b. \textit{Nature}, \textbf{509}, 63-65 (2014).

\item \label{KRK13} R. K. Kopparapu, \textit{et al.}, Habitable Zones around Main-sequence Stars: New Estimates. \textit{Astrophys. J.}, \textbf{765}, 131 (2013).

\item \label{BM98} A. F. C. Bridger, J. R. Murphy, Mars' surface pressure tides and their behavior during global dust storms. \textit{J. Geophys. Res.}, \textbf{103}, 8587-8601 (1998).

\item \label{AS10} P. Arras, A. Socrates, Thermal Tides in Fluid Extrasolar Planets. \textit{Astrophys. J.}, \textbf{714}, 1-12 (2010).

\item \label{LHE10} S. Lebonnois, \textit{et al.}, Superrotation of Venus' atmosphere analyzed with a full general circulation model. \textit{J. Geophys. Res. (Planets)}, \textbf{115}, 6006 (2010).

\item \label{Efr12} M. Efroimsky, Tidal dissipation compared to seismic dissipation: in small bodies, Earths, and super-Earths. \textit{Astrophys. J.}, \textbf{746}, 150 (2012).

\item \label{NL97} O. Neron de Surgy, J. Laskar, On the long term evolution of the spin of the Earth. \textit{Astron. Astrophys.}, \textbf{318}, 975-989 (1997).

\item \label{CLN03} A. C. Correia, J. Laskar, O. Neron de Surgy, Long-term evolution of the spin of Venus I. Theory. \textit{Icarus}, \textbf{163}, 1-23 (2003).

\item \label{KS90} S. Karato, H. A. Spetzler, Defect microdynamics in minerals and solid-state mechanisms of seismic wave attenuation and velocity dispersion in the mantle. \textit{Reviews of Geophysics}, \textbf{28}, 399-421 (1990).

\item \label{BRJ08} R. Barnes, S. N. Raymond, B. Jackson, R. Greenberg, Tides and the Evolution of Planetary Habitability. \textit{Astrobiology}, \textbf{8}, 557-568 (2008).

\item \label{ET08} L. T. Elkins-Tanton, Linked magma ocean solidification and atmospheric growth for Earth and Mars. \textit{Earth and Planetary Science Letters}, \textbf{271}, 181-191 (2008).

\item \label{HAG13} K. Hamano, Y. Abe , H. Genda, Emergence of two types of terrestrial planet on solidification of magma ocean. \textit{Nature}, \textbf{497}, 607-610 (2013).

\item \label{SCH12} S. M. Som, D. C. Catling, J. P. Harnmeijer, P. M. Polivka, R Buick, Air density 2.7 billion years ago limited to less than twice modern levels by fossil raindrop imprints. \textit{Nature}, \textbf{484}, 359-362 (2012).

\end{enumerate}

\subsection*{Acknowledgments}

J.L. thanks S. Lebonnois for providing the numerical outputs of the Venus model. The authors thank three anonymous referees for their insightful comments that significantly enhanced the manuscript. N.M. is supported in part by the Canada Research Chair program. This work was supported by grants from the Natural Sciences and Engineering Research Council (NSERC) of Canada to K.M. and N. M.

\subsection*{Supplementary Material}

\noindent Material and Methods

\noindent Supplementary Text

\noindent Figs. S1 to S5

\noindent References (25-35)



\clearpage


\clearpage

\begin{center}


\vspace*{3cm}
\subsection*{Supplementary Materials for}

\section*{Asynchronous rotation of Earth-mass planets in the habitable zone of lower-mass stars}
J\'er\'emy Leconte*, Hanbo Wu, Kristen Menou, Norman Murray

*Corresponding author. E-mail: jleconte@cita.utoronto.ca
\end{center}

\noindent \textbf{This PDF file includes:}

\indent Material and Methods\\
\indent Supplementary Text\\
\indent Figs. S1 to S5\\
\indent References (25-35)

\clearpage

\section*{Material and Methods}

We performed simulations of the thermal response of the atmosphere of terrestrial planets in the habitable zone for a wide range of forcing frequencies (solar day duration) using the LMDZ generic global climate model (\sect{sec:num_model}). In each of these simulations, the response is characterized in terms of the complex amplitude of the mass quadrupole ($\Pquad ( \omega-n,\ps,\Flux )$; (\sect{sec:num_torque})), which allows us to compute the atmospheric torque exerted by the star on the atmosphere. This model is tested against constraints from the Earth and Venus (\sect{sec:venus}).

We then show that, by analogy with the simple problem of a thermally forced, radiating slab, we are able to model analytically the frequency dependence of the atmospheric torque (\sect{sec:fit}). This allows us to provide an analytical formula that uses only two free parameters ($\amp(\ps,\Flux )$ and $\w0(\ps,\Flux )$) which can be calibrated from the simulations and are given in Table 1 for models bracketing the behavior expected in the habitable zone.

\section{Numerical atmospheric simulations}\label{sec:num_model}

To model thermal tides, we need to know the redistribution of mass in the atmosphere. As we will see, because of hydrostatic balance of the atmosphere, one only needs to know the spatial and temporal distribution of surface pressure. To model the evolution of the atmosphere, we use the Generic version of the LMDZ global climate model that has been specifically designed for exoplanet studies (\itref{LFC13}, \itref{FWM13}, \itref{LFC13b}). The details of the version used here can be found in ref.\,\ref{LFC13}.

For simplicity, we only model zero obliquity planets without topography, with a uniform surface albedo and thermal inertia (0.2 and 2000\,S.I. respectively), on circular orbits. The atmosphere is assumed to be dry and composed of an Earth-like mixture of N$_2$ with 376\,ppmv of CO$_2$, and radiative transfer is computed consistently using the correlated-k method. 

Despite the fact that the orbital distance of the habitable zone changes drastically for different stars, the flux defining its boundaries only changes by $\sim 25\%$ across the main sequence. Therefore, to reduce the computational cost, we ran suites of simulations for two insolations: i) an Earth-like case ($\Flux$\,=\,1366\,W.m$^{-2}$) which is representative of the inner edge, and ii) a $\Flux$\,=\,450\,W.m$^{-2}$ case which is representative of the outer edge. These two sets of simulations should encompass the behavior expected for planets in the habitable zone. As will be shown hereafter, however, the effect of the incoming stellar flux on the distance at which asynchronicity sets in is rather small compared to other effect such as the effect of mean surface pressure. 
For each insolation, several mean surface pressures (mainly 1 and 10\,bar) were used, and for each of these pressures, many simulations were done to cover the relevant range in forcing frequency $\omega-n$ (or equivalently the solar day duration, $2\pi/(\omega-n)$). This represents almost a hundred simulations. Snapshots of the pressure variations created by the diurnal tide in the \{1366\,W.m$^2$,\,10\,bar\} case for four different frequencies are shown in \fig{fig:diurnal_tides}.

As discussed in the main text, thanks to the carbonate-silicate cycle, CO$_2$ is thought to accumulate in the atmosphere near the outer edge to keep the surface unfrozen (\itref{KWR93}). Our $\Flux$\,=\,450\,W.m$^{-2}$ cases with a N$_2$ dominated atmosphere are indeed too cold for liquid water. We thus ran an additional suite of models for a 10\,bar, pure CO$_2$ atmosphere. Results show that changing from a N$_2$ to a CO$_2$ atmosphere reduces slightly the amplitude of the tide because of the increased albedo and higher absorption of CO$_2$ (see Table 1). This effect is however less important than the effect of the reduced flux.

As described hereafter, the amplitude and lag of the thermal tide is computed from each simulation (\sect{sec:num_torque}). Then, for a given set of parameters $\{\Flux,\ps\}$, the frequency dependence of the thermal tide is fitted following the method described in \sect{sec:fit}. Final results are given in Table 1.

\section{Torque due to thermal tides}\label{sec:num_torque}

Because of the non-spatially homogenous insolation that it receives, the atmosphere is continuously driven out of dynamical equilibrium and undergoes large-scale motions. This results in a non-spherically symmetric redistribution of its mass which thus can be gravitationally torqued by the star. To compute this torque, we first express the gravitational potential created by the atmosphere at a point $\vr\equiv \left( r,\theta,\phi\right) $ in space
\balign{
\PotTh(\vr)=-\G \int \frac{\d \mass'}{\left| \vr-\vrp\right|}=-\G \int \frac{ \rho(\vrp)}{\left| \vr-\vrp\right|} r'^2 \sin \theta' \d \theta' \d \phi' \d r', \nonumber
}
where $\G$ is the universal gravitational constant, $\rho(\vrp)$ is the density at the location $\vrp \equiv \left( r'\equiv \Rp+z,\theta',\phi'\right) $, and integration is carried out over the atmosphere. Assuming that the atmosphere is thin ($r'\sim \Rp$) and close to hydrostatic equilibrium, we have
\balign{
\frac{\d \press}{\d z}=-\grav \rho\ \Leftrightarrow\ \int_{z=0}^{\infty}\rho \d z = \int_{\press=0}^{\ps}\frac{ \d \press}{\grav}=\frac{\ps(\theta',\phi')}{\grav}, \nonumber
}
where $\press$ is the pressure, $\ps(\theta',\phi')$ its value at a point of the surface, and $\grav$ the surface gravity.

Expanding $\left| \vr-\vrp\right|$ in terms of Legendre polynomials ($\PLl$) and integrating over $r'$ we get
\balign{
\PotTh(\vr)=-\frac{\G \Rp}{\grav} \sum_{l=0}^{\infty} \left(\frac{\Rp}{r}\right)^{l+1}\int \PLl(\vr\cdot \vrp) \ps (\theta',\phi')\,\sin \theta' \d \theta' \d \phi'. \nonumber
}
Using the addition theorem, we expand $ \PLl(\vr\cdot \vrp)$ in spherical harmonics, which yields
\balign{
\PotTh(\vr)&=&-\frac{\G \Rp}{\grav} \sum_{l=0}^{\infty}\sum_{m=-l}^{l}  \frac{4\pi}{2 l+1} \left(\frac{\Rp}{r}\right)^{l+1}\Ylm(\theta,\phi)\int  \Ylmc(\theta',\phi') \ps(\theta',\phi') \,\sin \theta' \d \theta' \d \phi'  \nonumber\\
&=&-\frac{\G \Rp}{\grav} \sum_{l=0}^{\infty}\sum_{m=-l}^{l}  \frac{4\pi}{2 l+1} \left(\frac{\Rp}{r}\right)^{l+1} \plm\,  \Ylm(\theta,\phi),  \nonumber
} 
where $\plm\equiv\int \Ylmc\,\ps \,\sin \theta' \d \theta' \d \phi' $, are the complex moments of the pressure field.

The gravitational torque applied by the star on the atmosphere (the opposite of the torque by the atmosphere on the star) is thus given by $\mathbf{T}_\mathrm{a}=\Ms\,\vr\times\mathbf{\nabla} \PotTh$. For the sake of concreteness, we confine ourselves to the zero obliquity case where the torque is perpendicular to the planetary equator and is
\balign{\Ta=\Ms\,r\left(\frac{1}{r}\,\frac{\partial}{\partial \phi} \PotTh\right)=-\frac{\G \Ms\Rp}{\grav} \sum_{l=2}^{\infty} \frac{4\pi}{2 l+1} \left(\frac{\Rp}{r}\right)^{l+1}\sum_{m=-l}^{l} i \,m\,  \plm\,  \Ylm(\theta_\star=\frac{\pi}{2},\phi_\star),  \nonumber
}
where the dipole, $l=1$ term has been discarded as it is only a change in the center of mass (\itref{CL03b}). The colatitude and longitude of the substellar point ($\theta_\star$ and $\phi_\star$) are measured in the coordinate system defined by the spin vector and an arbitrary meridian of origin (the choice of the origin is inconsequential as long as the pressure moments are computed in the same coordinate system). Because, in our simulations, the amplitude of the $\plm$ decreases at higher $l$ and $\Rp/r\ll1$, we can keep the $l=2$ terms only. Furthermore, considering that $Y_2^{\pm 1}(\theta_\star=\frac{\pi}{2},\phi_\star)=0$ and $\tilde{p}_2^{-2}Y_2^{-2}=(-1)^{2\times2}\tilde{p}_2^{2*}Y_2^{2*}$, we have
\balign{\Ta&=&-\frac{\G \Ms\Rp}{\grav} \frac{8\pi}{5} \left(\frac{\Rp}{r}\right)^{3} i   \,Y_2^2(\frac{\pi}{2},0)\left( \ptwotwo \,e^{2i\phi_\star}-\pc e^{-2i\phi_\star}\right), \nonumber\\
&=&\sqrt{\frac{24\pi}{5}}\,\frac{\G \Ms\Rp }{\grav}  \left(\frac{\Rp}{r}\right)^{3}  \mathbb{I}\mathrm{m} (\Pquad),\nonumber\\
&=&\frac{3}{2}\,\Ka \left(\sqrt{\frac{10}{3\pi}}\, \mathbb{I}\mathrm{m} (\Pquad)\right),}
where $\mathbb{I}\mathrm{m}(\Pquad)\equiv\mathbb{I}\mathrm{m}( \ptwotwo e^{2i\phi_\star}) \equiv \left|  \ptwotwo\right| \sin (2\phi_\star-2\phia)$, where $2\phia$ is minus the argument of the complex number $\ptwotwo$. We also introduce $\dela\equiv\phi_\star-\phia$. Defining $\batm(2\freq)\equiv-\sqrt{\frac{10}{3\pi}}\ \mathbb{I}\mathrm{m}(\Pquad(\freq)) $, we recover the equation in the text. $\Pquad$ can be seen as the complex amplitude of the mass quadrupole in the reference frame where the substellar point is taken as the origin of longitudes. In a typical situation with no eccentricity, both $ \left|  \ptwotwo\right|$ and $\dela$ show only small fluctuations around a constant (see Fig.\,1). For higher rotation rates, baroclinic waves start to appear which increases the internal variability.

For illustrative purposes, we show the semi-diurnal component of the surface pressure field, i.e. $\ptwotwo Y_2^2+\tilde{p}_2^{-2}Y_2^{-2}$, in \fig{fig:diurnal_tides}. One can see that the pressure maximum lagging behind the star does so with a lag of more than 90$^\circ$ (in other words, the pressure maximum closest to the star is leading the motion of the substellar point) which is why atmospheric tides have an effect opposite to body tides. For these slow rotations, the diurnal component is still dominating the pressure distribution at the surface. This is the primary reason for the strong day/night side circulation seen in many 3D atmospheric models of tidally locked exoplanets (\itref{JHR93}, \itref{YCA13}, \itref{LFC13}), but the amplitude of this component becomes smaller at Earth-like rotation rates (see below).

\section{Application to Earth, Venus, and more massive planets}\label{sec:venus}

Our most extensive source of data on thermal tides coming from the Earth, we have validated our numerical model by running a simulation for this specific case (accounting for the actual topography and albedo variations of the planet; see ref.\,\ref{LFC13b}).
To compare with meteorological data, we compute the amplitude and phase of both the diurnal and semi-diurnal component of the tidal pressure wave at each surface grid point. This is done by performing a Fourier analysis of the pressure time series output by the simulation (\itref{CL70}).
In this case, the spatial distribution of the semi-diurnal phase and amplitude predicted by our atmospheric model is fairly close to the data (see \fig{fig:earth_comp} to be compared with Fig. 2S.3 of ref.\,\ref{CL70}). Our semi-diurnal solar tide has a $\sim$2\,mb amplitude and $\approx150^{\circ}$ phase at the equator which is in good agreement with the meteorological observations ($\sim$1.2\,mb and $\approx150-160^{\circ}$; \itref{CL70})
We also recover the fact that, for a fast rotating planet like the Earth, the diurnal component of the tides has a small amplitude compared to the semi-diurnal one

Note, however, that for the Earth, both gravitational and atmospheric torques (created by the Sun) are too weak to bring the Earth to an equilibrium spin state over its lifetime. This should be true for planets in the habitable zone of stars above $\sim0.9\msun$. These objects are thus not expected to be found in an equilibrium spin state (either synchronous or not).
We also tested the model on Mars where the amplitude of the pressure variations at the surface were found to be in agreement with surface data (\itref{BM98}). We do not show these results here because the Martian atmosphere is too thin ($\sim6$\,mb) to significantly affect the spin of the planet.

Because the amplitude of the thermal tide is greater in a 10\,bar atmosphere than in a 1\,bar atmosphere, one might expect that the tide would be stronger in a Venus-like atmosphere, because it is even more massive.
There is, however, no contradiction with our theory as we find that the amplitude of thermal tides does \textit{not} increase continuously with surface pressure. Tests show that above a surface pressure of a few tens of bars (depending on the atmospheric composition), the amplitude of thermal tides starts decreasing again. This is caused by stellar energy being progressively absorbed higher in the atmosphere and the tide being damped by a growing quiescent layer beneath it. For very massive atmospheres, the system should slowly transition toward a regime where the solid surface has a negligible effect, although the thermal tide itself might still impact the spin of the planet (\itref{AS10}). The exact mass of atmosphere at which this transition occurs needs to be investigated in more detail. 

Venus, in addition to its thick atmosphere, exhibits very reflective and absorbing clouds that strongly decrease the amount of light absorbed in the lower atmosphere, and thus the forcing. Indeed, only a few watts per square meters reach the surface. To take that into account, we use outputs of the Venus version of the LMDZ model (\itref{LHE10}) and compute the amplitude of thermal tides as described above. 
This yields an amplitude $|\Pquad|\sim$1-1.6\,mb for the semi-diurnal solar tide (as shown in Fig.\,1.B), and a torque that is roughly equal to the gravitational torque expected for Venus (\itref{DI80}) for a quality factor of $Q\sim100$, which is in the range of what can be expected for a rocky mantle without oceans. This shows that, even in this well constrained case, our framework yields sensible results. To facilitate comparisons, this value of the quality factor for bodily tides is the fiducial value that we use when needed, but we also show the sensitivity of our results on this parameter that may vary from a planet to another (especially in Fig.\,3). The values of $\amp$ and $\w0$ needed to reproduce $\Pquad$ are given in Table 1.

\section{A thermal inertia model for the atmospheric bulge}\label{sec:fit}

In first approximation, the atmospheric bulge created by the periodic insolation pattern is not aligned with it only because it takes some time for the atmosphere and surface to increase their temperature when heated. Thermal inertia of the atmosphere-surface system should thus be an important parameter in determining both the amplitude and the lag of the thermal tide. 

This problem should be conceptually analogous to the classical problem of a periodically heated slab (\itref{ID78}). With that in mind we derive the thermal response of an idealized slab and will later show that the frequency dependence of the solution captures, rather surprisingly, striking features of the thermal tide response.

Let us consider a slab of heat capacity $\Cap$ per unit area at temperature $T$, radiating like a black body, and receiving an energy flux $F=F_0+\dF$. Let us assume that the temperature response can be linearized around a mean temperature, $T=T_0+\delta T $. The temperature evolution can thus be written
\balign{\label{eqslab}
\Cap\frac{\d T}{\d t}=F-\ssb T^4 \  \Leftrightarrow\ \Cap\frac{\d }{\d t}\delta T=F_0-\ssb T_0^4+\dF - 4 \ssb T_0^3 \delta T.
}
Now, if we assume that the mean temperature is such that the slab is in mean thermal equilibrium ($F_0=\ssb T_0^4$), and that the perturbation is periodic of the form $e^{i\freq t}$ (because we consider only one component of the Fourier decomposition of the insolation, e.g. the semi-diurnal one), the complex amplitude of the response is given by
\balign{i \freq \Cap\dT = \dF - 4 \ssb T_0^3 \dT\  \Leftrightarrow\ \dT= \frac{\dF}{2\w0\Cap}\frac{1}{1+i\freq/(2\w0)},}
with $\w0\equiv 2 \ssb T_0^3/C$, which is twice the inverse of the usual thermal timescale. Interestingly, if the heat capacity of the atmosphere/surface system to be modeled can be estimated (for example by $\Cap=\cp\ps/\grav+C_\mathrm{surf}$), one can have an order of magnitude estimate for $\w0$,
\balign{\w0=\frac{ 2 \ssb T_0^3}{\cp\ps/\grav+C_\mathrm{surf}}.}
Note that $T_0$ is the emission, and not surface, temperature.
This order of magnitude estimate is actually confirmed by the numerical model. If further scalings are needed, let us note that $\w0\propto T_0^3\propto \Flux^{3/4}$. For a given star we can further say that $\w0\propto a^{-3/2}\propto n$.

By analogy with our problem where the important frequency is $\sigma=2\omega-2n$, let us model the pressure bulge amplitude and lag by
\balign{\Pquad= -\frac{\amp}{1+i(\omega-n)/\w0}\ \Leftrightarrow\ \mathbb{I}\mathrm{m}(\Pquad)= \amp\,\frac{(\omega-n)/\w0}{1+\left[(\omega-n)/\w0\right]^2},}
where $\amp(\ps,F_0)$ and $\w0(\ps,F_0)$ can be fitted from a set of numerical atmosphere models. Despite its simplicity, and the lack of proper dynamical treatment, Fig.\,1 shows that this parametrization of the torque actually captures salient features of the complex frequency dependence of the atmospheric response, especially when considering the huge uncertainties in other parameters.
Then, thanks to the simplicity of our approach, we can describe the thermal response of a given atmosphere with only two parameters that can be computed for a given set of atmospheric parameters (mean pressure, incoming flux, etc.). Results for our various sets of simulations encompassing the habitable zone can be found in Table 1.

Giving an analytical order of magnitude estimate of the amplitude of the disturbance (especially in terms of pressure) is however rather difficult because one would need to solve both dynamical and radiative transfer equations. The important, and counter-intuitive, differences between our results for the Earth and Venus show that more work is needed to find a reliable analytical estimate. For the time being, the model thus still needs to be calibrated on numerical results as the one presented in Table 1.

The main difference with earlier analytical treatments (\itref{DI80}, \itref{CL01}) is that we account for the cooling feedback of the atmosphere (last term in Eq.\,(2)) which yields a response $\propto 1/(\w0+i \freq)$ instead of $1/i \freq$. The former formula was found in ref.\,\ref{ID78}, but their equivalent of $\w0$ was unknown and assumed to be zero, yielding the later formula. Our treatment thus avoids any discontinuity in the zero frequency limit and self-consistently predicts a linearly increasing, then saturating, lag angle.

\section{Existence and calculation of asynchronous equilibrium spin states for various rheological models}

As shown in ref.\,\itref{CL01}, independently of the tidal models used, finding equilibrium spin states reduces to solving the following equation
\balign{
f(\omega-n)\equiv\frac{\batm(2\omega-2n)}{\bgrav(2\omega-2n)}=-\frac{\Kg}{\Ka}.}
Here we take the convention that the obliquity is always 0$^\circ$ and that $\omega$ is negative for a retrograde rotation, but we keep in mind that for each equilibrium state $(\omega,0^\circ)$ found, the $(-\omega,180^\circ)$ state is also an equilibrium (\itref{CL01}).

The location and number of solutions, however, does depend on the specific model used. If the restriction $\tilde{f}$ of $f$ to the positive frequency range is monotonic around possible equilibria, only one synchronous and two asynchronous spin states exist (\itref{CL01}). An important qualitative novelty of the present work is that $\batm$ is not monotonic. As a result, more equilibria potentially exist, depending on the rheology used to describe the bulk of the planet.
In general, this rheology is parametrized by a frequency dependent quality factor $Q(\freq)$ with
\balign{\bgrav(\freq)=\mathrm{sign}(\freq)\frac{k_2(|\freq|)}{Q(|\freq|)},}
where $k_2$ is the Love number of degree 2.

Recent studies suggest that tidal dissipation inside terrestrial planets can be accurately described by the so called Andrade rheology which accounts for both the viscous behavior of the mantle in the zero frequency limit (very close to synchronization) and the low frequency dependence at higher frequencies (similar to the constant-$Q$ model). Thus, we first discuss the existence of equilibrium asynchronous spin states with the Andrade model. Because a simple analytical solution is difficult to attain, we then show how the qualitative behavior of the solution can be understood in terms of two simpler models (the constant time lag and constant-$Q$ models) that allow us to find analytical solutions in limit cases.

We do not consider the viscoelastic Maxwell model (which disregards inelastic behaviors which are important in the body of the planet) because its frequency dependence at higher frequencies ($k_2/Q\propto 1/\freq$) is to steep to properly explain existing data for terrestrial planets (\itref{Efr12}).
Indeed, the Earth's $Q$ changes by slightly more than an order of magnitude between the Chandler wobble period (about 440 days) and seismic periods of a few seconds (\itref{Efr12}, \itref{CLN03}, \itref{KS90}) whereas the Maxwell model would predict a change of about 7 orders of magnitude ($|440 \,\mathrm{days}/1\,\mathrm{sec}|\approx4\times10^{7}$).

\subsection{The Andrade model}

Although viscosity dominates the dissipation in the mantle at very small frequencies ($\freq\ll \tauM^{-1}$, where $\tauM$ is the Maxwell time), inelasticity takes over at higher frequencies. Both these behaviors are encompassed in the Andrade model (see ref.\,\itref{Efr12} for details) whose dissipation is described by
\balign{\nonumber
\frac{k_2(\freq)}{Q(\freq)}=\frac{3}{2} \Al\frac{\freq \left(\tauM^{-1}+\freq^{1-\aM}\tauM^{-\aM}\sin (\frac{\aM \pi}{2})\,\Gamma(1+\aM)\right)}{\left((1+\Al)\freq+\freq^{1-\aM}\tauM^{-\aM}\cos (\frac{\aM \pi}{2})\,\Gamma(1+\aM)\right)^2+\left(\tauM^{-1}+\freq^{1-\aM}\tauM^{-\aM}\sin (\frac{\aM \pi}{2})\,\Gamma(1+\aM)\right)^2},
}
where $\Al$ is a constant of order unity depending on the mass, radius and rigidity of the planet, and $\Gamma$ is the gamma function. For the Earth, the Maxwell time, $\tauM$, is about 500\,yr, and the index of the frequency power law at higher frequencies, $\aM$, is estimated between 0.15 and 0.3 (we will use 0.2 as a fiducial value). 

As is visible in Fig.\,2, when the amplitude of the thermal tide reaches a threshold, there are five different equilibria. This is due to the fact that $\tilde{f}$ is non monotonic over the range of frequencies of interest. However, to give an analytical formula for this threshold we need to simplify the problem further.

\subsection{The constant-$Q$ model}

An important simplification comes from the fact that the Maxwell time for terrestrial planet is much larger than the forcing periods we are envisioning. Therefore, for forcing periods of the order of days to months ($\freq\gg \tauM^{-1}$), the tidal dissipation scales as $\freq^{-\aM}$ which is weakly frequency dependent ($\aM\sim0.2$).
This is needed to explain the low frequency dependence of the Earth $Q$ (see above) which stems from the low frequency dependence of inelastic processes in rocky mantles.
Originally, it is why it has very often been assumed that $Q(\freq)$ is independent of frequency for rocky planets, giving rise to the so-called constant-$Q$ (or constant phase lag) model.

Indeed, as seen in Fig.\,2, this very simple model reproduces most of the features of the full Andrade model. In particular, we recover the existence of four asynchronous equilibria when a given criterion is met. This criterion is given by 
\balign{\label{gammaQ}
\gamma_{Q}\equiv \sqrt{\frac{5}{6\pi}}\,\frac{\Ka}{\Kg}\frac{\amp Q}{k_2}\ge 1.
}
We can also calculate the location of the various spin states
 \balign{\label{asyncQ}
 \omega_{\pm}^\epsilon=n\pm\w0\left( \gamma_{Q}+\epsilon\sqrt{\gamma_{Q}^2-1}\right),}
 where $\epsilon=1$ for stable equilibria and $\epsilon=-1$ for unstable ones. This equation can be deduced from Eq. 7 in ref.\,\itref{CL01} applied to our specific rheology. It is noteworthy that, because of the shape of the various torques, the degree of asynchronism ($|\omega_{\pm}-n|$) does not continuously goes to zero (as can be seen in \fig{fig:asynchronicity}), and is always greater than $\w0$.
 
 Because it is completely analytical, and considering the small differences with the full Andrade model compared to the uncertainties in the various parameters involved, we mainly discuss the results of this model in the body of the article. 

The main limitation of this model is that it predicts a discontinuity near synchronization where the frequency vanishes. It thus cannot be used to describe what happens to the synchronous spin state. Various authors have tried to overcome this problem by changing $Q$ near $\freq=0$ so as to have a linear behavior but the range of frequency over which they made the transition was mostly unconstrained. This is resolved in the Andrade model. 


\subsection{The viscous model}

Indeed, in the zero frequency limit, the Andrade model tells us that the mantle acts as a viscous fluid, for which the time delay between the perturbation and the response is constant so that $Q^{-1}\approx\dt\freq$. In this regime, $f=- \sqrt{\frac{5}{6\pi}}\,\frac{\amp}{\k2\dt\w0} \frac{1}{1+(\freq/\w0)^2}$. For this specific model, $\tilde{f}$ being monotonic, there is at most two asynchronous spin states which exist \textit{only} when the following criterion is met
\balign{\label{viscouscriterion}
\gamma_{\mathrm{vis}}\equiv \sqrt{\frac{5}{24\pi}}\,\frac{\Ka}{\Kg}\frac{\amp}{\k2\dt\w0}\ge 1.
}
Then, the two equilibrium rotation rates are given by $\omega_{\pm}=n\pm\w0 \sqrt{\gamma_{\mathrm{vis}}-1}.$
It can be shown that this criterion is also sufficient for the synchronous spin state to be an unstable equilibrium.


 A simple estimation for Venus yields $\gamma_{\mathrm{vis}}<10^{-7}.$ This is because for a rocky, Earth-like mantle, $\tauM$ is extremely large ($\sim500$\,yr), and $\k2\dt=3/2\times\Al\tauM$ in this context. As a result, for a body whose rheology obeys the Andrade law, it is very likely that the synchronous spin state is a stable equilibrium. This has important implications as discussed below.
 
For sake of completeness, note that, if the planet were to behave as a viscous fluid over the whole frequency domain with a much lower viscosity, then the situation might be a little different. Calibrating the time lag on the Earth Moon system, one finds $\dt\sim630\,s\ll\tauM$ (\itref{NL97}). Then, criterion (\itref{viscouscriterion}) would be much easier to fulfill, and the planet could have two stable, asynchronous equilibria and one unstable synchronous one, as depicted in \fig{fig:torque2}. Although often used for its linear behavior, data showing the inadequacy of this model to describe solid planets are accumulating (\itref{Efr12}). We thus mention this only to help the reader make the link with the various models available in the literature. 


\subsection{Critical distance for the existence of asynchronous equilibrium spin states}

Rewriting the criterion given by \eq{gammaQ}, one can see that planets are expected to have non-synchronous, equilibrium spin states when located beyond a critical orbital distance given by 
\balign{
 \ac=\left(\frac{10\pi}{3}\right)^{1/6}\left(G  \Ms  \rhob\Rp^2\,\frac{\k2}{\amp Q}\right)^{1/3}=\left(\frac{15}{8\pi}\right)^{1/6}\left(\grav  \Rp\Ms\frac{\k2}{\amp Q}\right)^{1/3}.
}
which is plotted in Fig.\,3 for an Earth-like planet with $\ps=1$ and 10\,bar and the value of $\amp$ corresponding to $\Flux=1366\,$W.m$^{-2}$.
The only dependence in the flux received by the planet appears through $\amp$. However, although $\amp$ increases with flux as expected, it changes by less than $\sim25\%$ across the habitable zone despite a flux variation of a factor 3 (see Table 1). As a result, as shown in \fig{fig:Fdependence}, for a given pressure, the explicit dependence of the critical distance on the flux received by the planet is small compared to the pressure dependence (and the uncertainty on the bodily tides). For planets outside the habitable zone, receiving much larger/lower fluxes, additional simulations would be needed.

\section{Stellar properties and habitable zone}

As discussed in the main text, we often have to relate the flux received by a given planet, its orbital distance and the stellar mass. For example, in \fig{fig:asynchronicity}, once the atmospheric properties and the substellar top-of-atmosphere insolation $\Flux$ are chosen, the orbital distance will vary with the stellar mass. To do so, the bolometric luminosity ($\Ls$) of a star of $\Ms$ is given by (\itref{BRJ08})
\balign{\Ls=\lsun\, 10^{4.101 \mu^3+8.162 \mu^2+7.108 \mu},}
where $\mu\equiv\log (\Ms/\msun)$. Then the distance at which a planet receives $\Flux$ is given by
$a=\sqrt{\frac{\Ls}{4\pi\Flux}}$. 

For the boundaries of the habitable zone shown in Fig.\,3 and used throughout the text, we used the parametrization of the outer edge from ref.\,\ref{KRK13}. The parametrization of the inner edge is taken from the same reference but is rescaled to take into account recent results from 3D climate models (\itref{LFC13b}); i.e. the dependence on the effective temperature of the star is left unchanged, but the inner edge is set at 0.95\,AU around a Sun-like star.

\clearpage

\section{Trapping in synchronous resonance and constraints on atmospheric evolution}

If a solid planet with a non-zero triaxiality ever approaches near the 1:1 synchronous spin-orbit resonance, the conservative restoration torque exerted by the star on its permanent asymmetry can, in some cases, trap the planet in spin-orbit resonance. As pointed out in ref.\,\itref{CL01} and \itref{CLN03}, even if the synchronous rotation rate were made unstable by thermal tides, spin-orbit trapping could, in some cases, stabilize a planet in this state if it ever went near it (if the atmosphere was formed a long time after the planet's formation, for example).

However, our more realistic treatment of both thermal and body tides has made this issue rather moot. Indeed, with a realistic rheology, the synchronous spin state is found to be stable (see Fig.\,2) so that a planet whose spin approaches this region will end-up being synchronized even without spin-orbit trapping. 

But we know that Venus did \textit{not} get caught in a synchronous spin state while having an internal structure quite similar to the Earth (which is well modeled by an Andrade rheology). So this shows that planets can, and some will, avoid evolving toward such a stable synchronous state. 

This creates a very interesting, new constraint. With our treatment, even without resonance trapping, if Venus spins down fast enough to have a rotation rate between the two unstable equilibria in Fig.\,2 (which could act as repelling barriers) before the atmosphere is created, then even once the atmosphere is outgassed, the spin-down will probably proceed toward synchronization. This tells us that the atmospheric build-up on Venus had to occur on a shorter timescale than the tidal despinning, which constrains both the dissipation efficiency in the planet and its atmospheric evolution. 

Fortunately, this is in good agreement with the recent developments in the theory of atmosphere formation in general, and on Venus in particular. These developments suggest that just after the formation of the planet itself, the mantle is in a magma ocean state and the atmosphere consist of tens of bars of CO$_2$ and water in solution equilibrium with the magma. Then the magma ocean solidifies (after one to tens of Myr) and hundreds of bars of CO$_2$ (up to thousands for H$_2$O) are released to the atmosphere because these gases are poorly soluble in solid rocks (\itref{ET08}).

After that, for Venus, ref.\,\itref{HAG13} suggests that the atmosphere never cooled down sufficiently to allow the water to condense and that the water just escaped to space in the first tens to hundreds Myr living a thick CO$_2$ atmosphere behind. Thus, even if this atmosphere subsequently evolved, it is unlikely that it ever got orders of magnitude thiner than the present one. The atmospheric tide may thus have been efficient during most of Venus' history. 

For the Earth, water is expected to have condensed when the surface cooled enough, creating the oceans and reducing the mass of the atmosphere to a value that is commensurate with the current one. Then, the longer, secondary, atmospheric evolution began. With that in mind, it is unlikely that the Earth ever had a much thinner atmosphere than its present one. Some data even indicate that it could have been thicker during some periods in its evolution (\itref{SCH12}). Again, this supports the idea that thermal tides may have been efficient for most of the history of planets in the habitable zone who could retain a significant atmosphere.

\clearpage
\setcounter{figure}{0}

\renewcommand{\thefigure}{S\arabic{figure}}

\begin{figure}[htbp] 
 \centering
\subfigure{ \includegraphics[scale=1.,trim = 0cm 1.cm 0.cm 0.cm, clip]{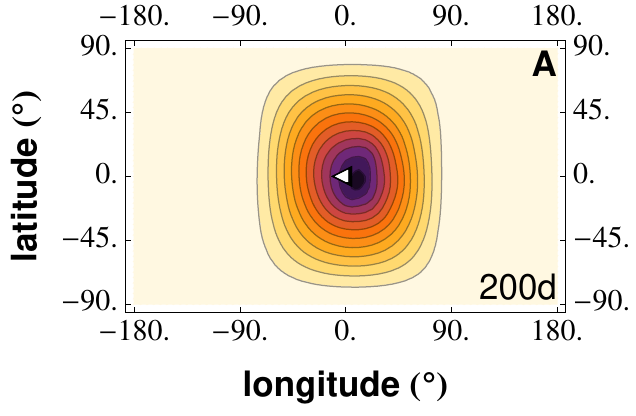} }
\subfigure{ \includegraphics[scale=1.,trim = .8cm 1.cm 0.cm 0.cm, clip]{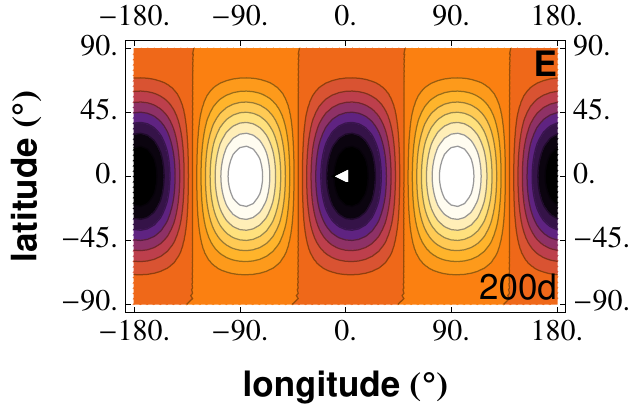} }\\
\vspace{-0.3cm}
\subfigure{ \includegraphics[scale=1.,trim = 0cm 1.cm .0cm .3cm, clip]{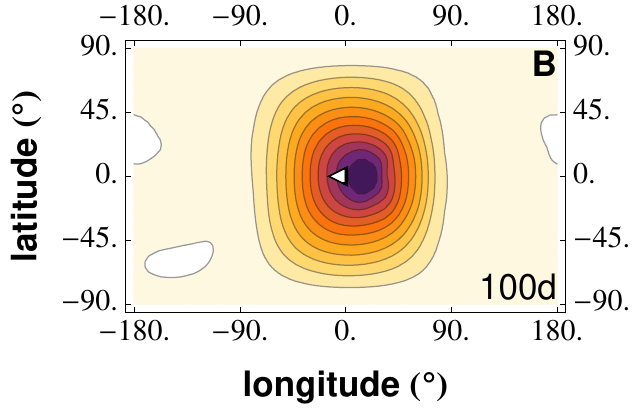} }
\subfigure{ \includegraphics[scale=1.,trim = .8cm 1.cm 0.cm .3cm, clip]{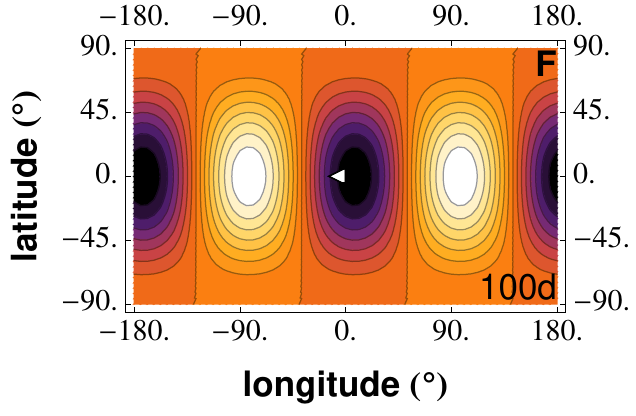} }\\
\vspace{-0.3cm}
\subfigure{ \includegraphics[scale=1.,trim = 0cm 1.cm .0cm .3cm, clip]{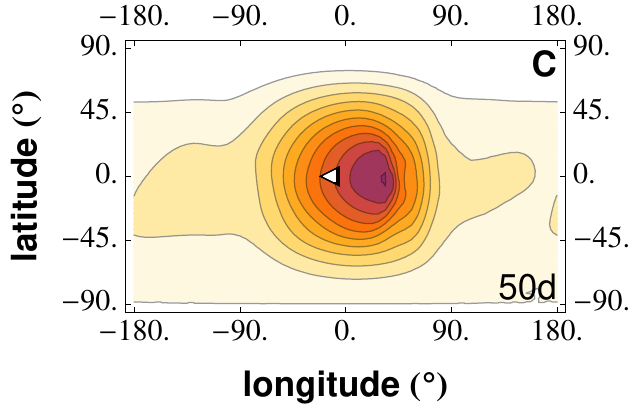} }
\subfigure{ \includegraphics[scale=1.,trim = .8cm 1.cm 0.cm .3cm, clip]{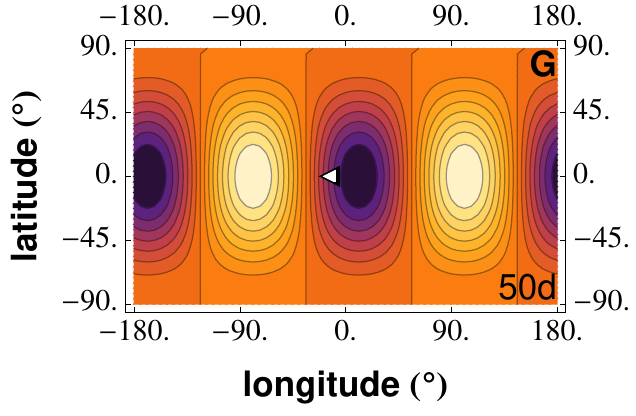} }
\vspace{-0.3cm}
\subfigure{ \includegraphics[scale=1.,trim = 0cm 0.cm .0cm .3cm, clip]{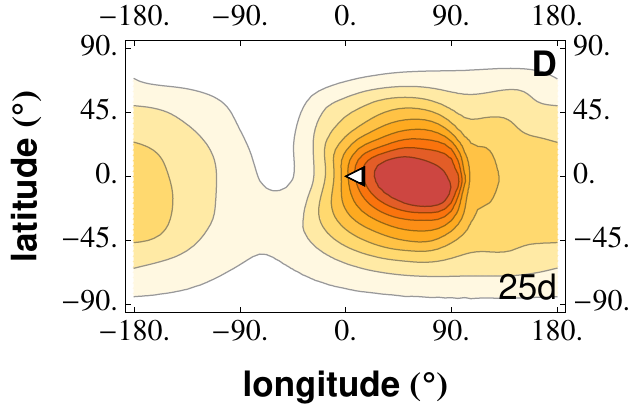} }
\subfigure{ \includegraphics[scale=1.,trim = .8cm 0.cm 0.cm .3cm, clip]{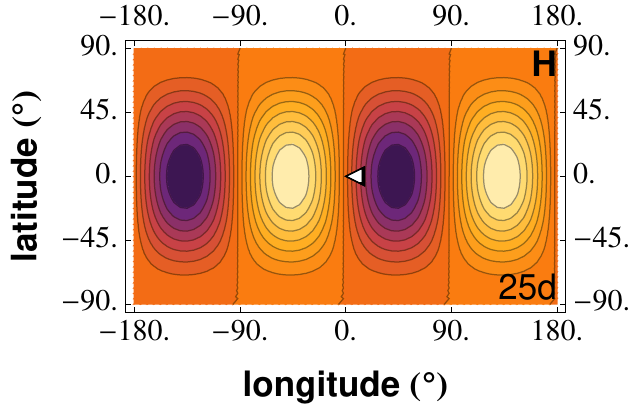} }\\
\subfigure{ \includegraphics[scale=.8,trim = 0cm 0.cm .0cm .cm, clip]{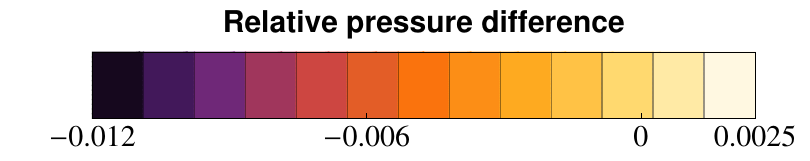} }
\subfigure{ \includegraphics[scale=.8,trim = 0cm 0.cm .0cm .cm, clip]{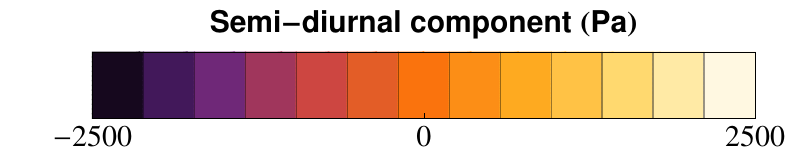} }\\
\caption{
\textbf{Surface pressure anomaly created by the thermal tide.} \textbf{A}-\textbf{D}: Snapshots of the spatial distribution of the departure of the surface pressure from its mean value created by the thermal tide. \textbf{E}-\textbf{H}: Spatial distribution of the semi-diurnal component only.
The location and direction of motion of the substellar point is shown with a white arrow. From top to bottom, the length of the solar day is decreased from 200 (planet near synchronization) to 25\,days for a model with a 10\,bar mean surface pressure on a 100\,d orbit. This shows the increase in the lag and decrease in amplitude with increasing forcing frequency.
}
 \label{fig:diurnal_tides}
\end{figure}

\begin{figure}[htbp] 
 \centering
\subfigure{ \includegraphics[scale=1.2,trim = 0cm .95cm 0.cm 0.cm, clip]{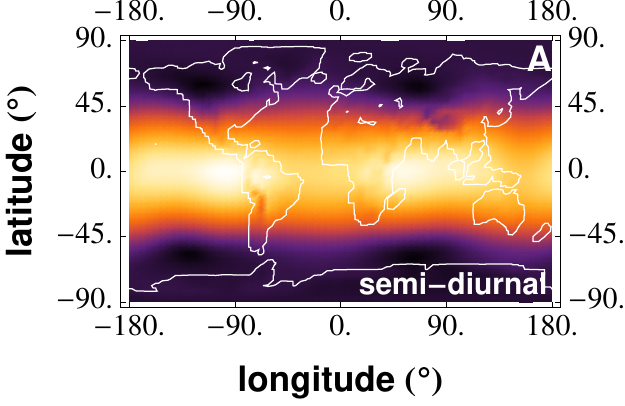} }\\
\subfigure{ \includegraphics[scale=1.2,trim = 0cm .cm .0cm .27cm, clip]{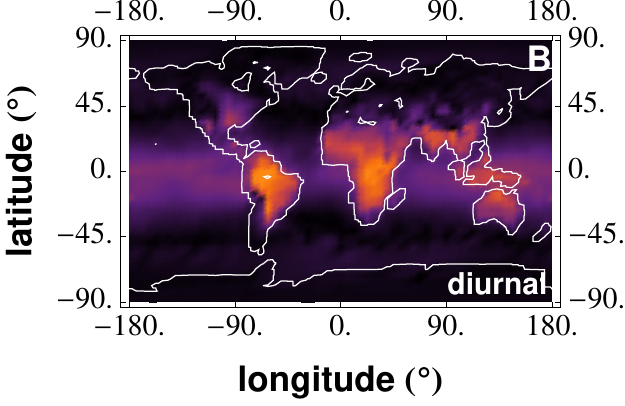} }\\
\subfigure{ \includegraphics[scale=.8,trim = 0cm .cm .0cm .cm, clip]{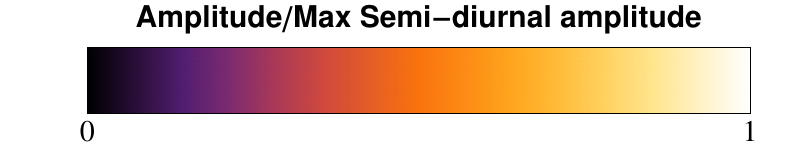} }\\
\subfigure{ \includegraphics[scale=1.2,trim = 0cm 0.cm .0cm .3cm, clip]{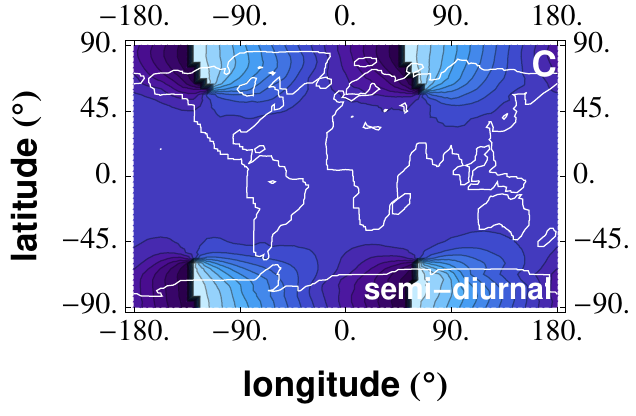} }\\
\subfigure{ \includegraphics[scale=.8,trim = 0cm 0.cm .0cm .cm, clip]{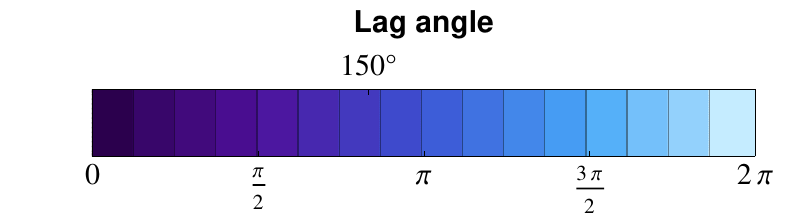} }\\
\caption{
\textbf{Amplitude and phase of the thermal tide for the Earth model.} \textbf{A}: Map of the amplitude of the semi-diurnal component of the surface pressure tide (normalized to its maximum value) for the Earth validation model. \textbf{B}: Amplitude of the diurnal tide normalized to the maximum amplitude of the semi-diurnal component. This shows that, on Earth, the semi-diurnal component is dominating. \textbf{C}: Phase of the semi-diurnal tide. The average value at the equator is $\approx150^{\circ}$, close to the expected value. To be compared with Fig. 2S.3 of ref.\,\ref{CL70}.
}
 \label{fig:earth_comp}
\end{figure}

\begin{figure}[htb] 
 \centering
\subfigure{ \includegraphics[scale=1.5,trim = 0cm 0.cm 0.cm 0.cm, clip]{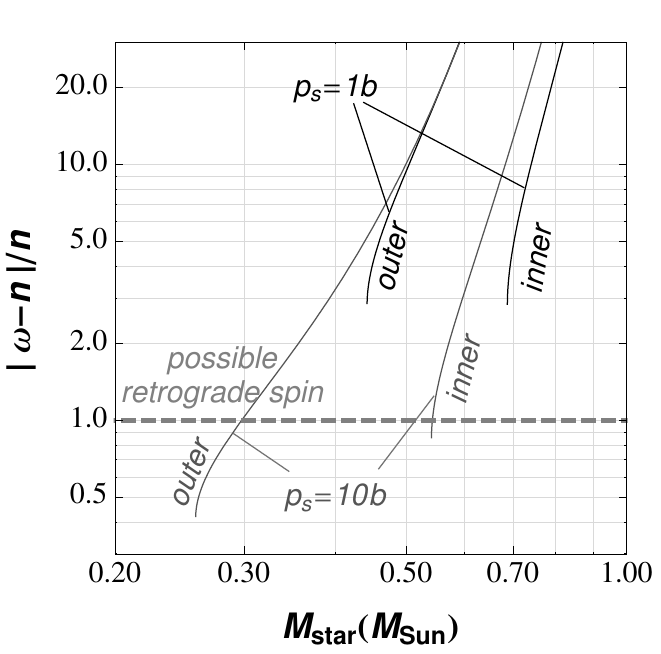} }
\caption{
\textbf{Degree of asynchronicity as a function of stellar mass for Earth-like planets in the habitable zone.} $|\omega_\pm-n|/n$ is computed from \eq{asyncQ} for planets with different atmospheric masses (characterized by the surface pressure, $\ps$), near both the inner and the outer edge of the habitable zone. We only show stable equilibria.
}
 \label{fig:asynchronicity}
\end{figure}

\begin{figure}[p] 
 \centering
\subfigure{ \includegraphics[scale=1.,trim = 0cm .9cm 0.cm 0.cm, clip]{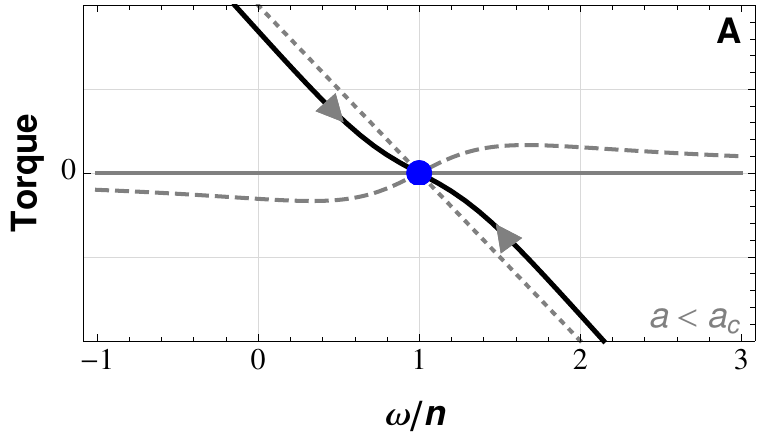} }\\
\vspace{-0.4cm}
\subfigure{ \includegraphics[scale=1.,trim = 0cm .9cm .0cm 0cm, clip]{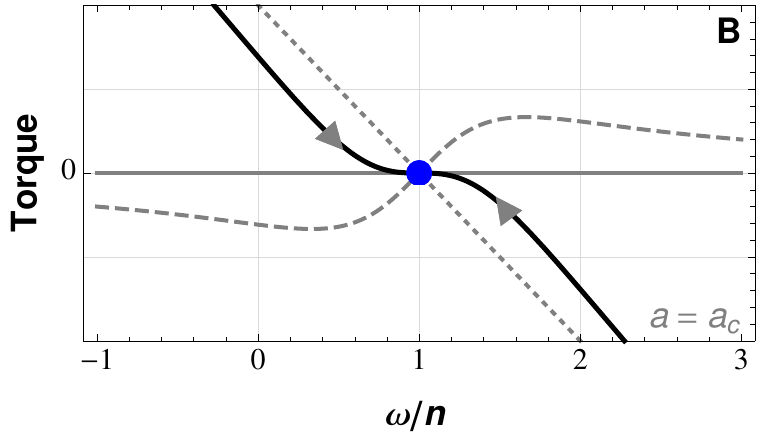} }\\
\vspace{-0.4cm}
\subfigure{ \includegraphics[scale=1.,trim = 0cm .cm .0cm 0cm, clip]{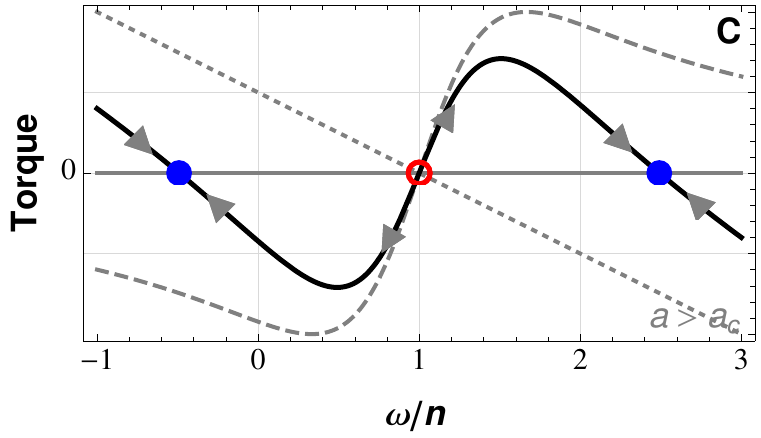} }
\caption{
\textbf{Equilibrium spin states of the planet in the weak friction approximation.} Atmospheric (dashed), gravitational (dotted) and total (solid) torque as a function of spin rate for the viscous model. Arrows show the sense of spin evolution. \textbf{A}: weak atmospheric torque, only one equilibrium, synchronous spin state exists (blue disk). \textbf{B}: bifurcation point ($a=\ac$). \textbf{C}: the atmospheric torque is strong enough to generate two \textit{stable}, asynchronous, equilibrium spin states (blue disks; one is retrograde in the case shown). The synchronous spin state is \textit{unstable}.
}
 \label{fig:torque2}
\end{figure}

\begin{figure}[htb] 
 \centering
\subfigure{ \includegraphics[scale=1.4,trim = 0cm 0.cm 0.cm 0.cm, clip]{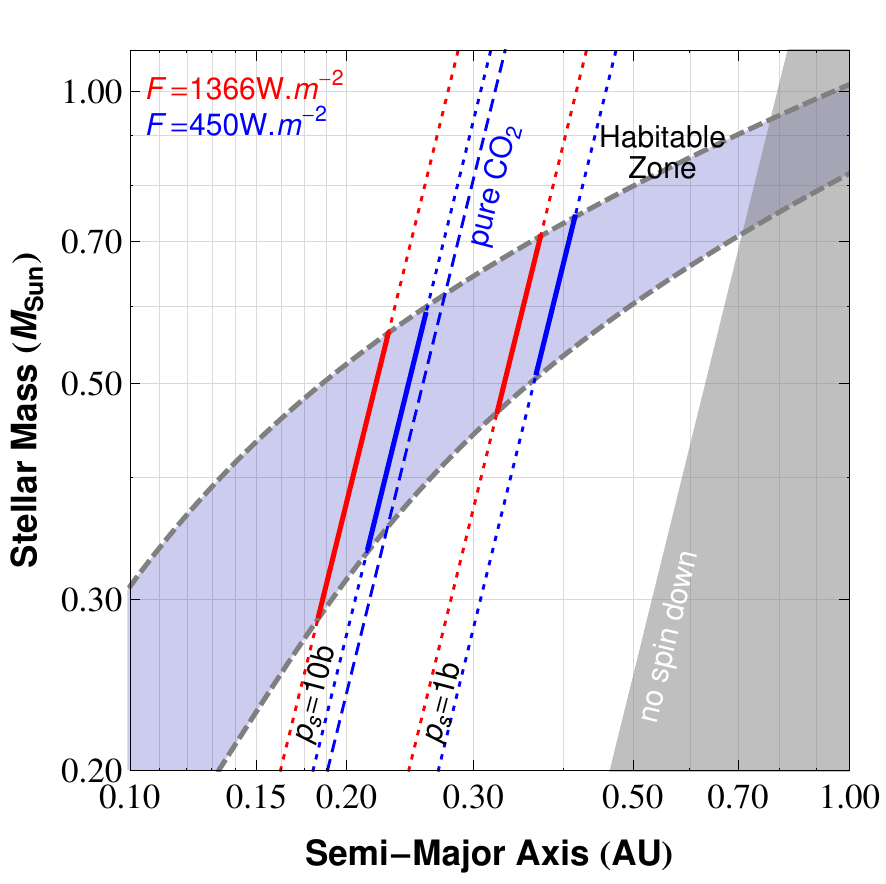} }
\caption{
\textbf{Dependence of critical "non-synchronous" orbital distance on planetary flux.}
For each pressure ($\ps=1$ and 10\,bar), the critical distance, $\ac$, beyond which planets are expected to exhibit non-synchronous equilibrium spin states is shown for two extreme values of $\amp$ representative of the conditions in the inner (red) and outer (blue) regions of the habitable zone. The dashed line corresponds to the pure CO$_2$ outer edge case. Other features are defined in Fig.\,3. For a given atmospheric pressure, the critical distance varies little with the value of the flux received by the planet in the atmospheric model, especially when compared to the variations with pressure.
}
 \label{fig:Fdependence}
\end{figure}

\end{document}